\documentclass[twocolumn, notitlepage, aps, prl, superscriptaddress, nofootinbib]{revtex4-1}
\usepackage[]{sidecap}
\usepackage[]{amsmath}
\usepackage{commath}
\usepackage{amssymb}
\usepackage{amsfonts}
\usepackage[usenames,dvipsnames,svgnames,table]{xcolor}
\usepackage{graphicx}
\usepackage{natbib}
\usepackage{IEEEtrantools}
\usepackage[]{hyperref}

\newcommand*\oline[1]{%
  \vbox{%
    \hrule height 0.4pt
    \kern0.35ex
    \hbox{%
      \kern-0.1em
      \ifmmode#1\else\ensuremath{#1}\fi
      \kern-0.1em
    }
  }
}

\begin{document}

\baselineskip 15pt

\title{Optimal Transport in Worldwide Metro Networks}

\author{\small Wei Li,$^{1,2}$ Jiao Gu,$^{2,*}$ Shiping Liu,$^{3,2}$ Yueying Zhu,$^1$ \\
Shengfeng Deng,$^1$ Longfeng Zhao,$^1$ Jihui Han,$^1$ and Xu Cai$^1$\\[3pt]
\it
$^1$College of Physical Science and Technology, Huazhong Normal
University, Wuhan 430079, China.\\
$^2$Max-Planck-Institute for Mathematics in the Sciences, Inselstr.22,
04103 Leipzig, Germany.\\
$^3$Department of Mathematical Sciences, Durham University, \\
Science Laboratories South Rd, Durham DH1 3LE, UK.\\
$^*$To whom correspondence should be addressed. Email: jiao.gu@mis.mpg.de
}

\begin{abstract}

Metro networks serve as good examples of traffic systems for understanding the
relations between geometric structures and transport properties. We study and
compare 28 world major metro networks in terms of the Wasserstein distance, the
key metric for optimal transport, and measures geometry related, e.g.~fractal
dimension, graph energy and graph spectral distance. The finding of power-law
relationships between rescaled graph energy and fractal dimension for both
unweighted and weighted metro networks indicates the energy costs per unit area
are lower for higher dimensioned metros. In $L$ space, the mean Wasserstein
distance between any pair of connected stations is proportional to the fractal
dimension, which is in the vicinity of our theoretical calculations treated on
special regular tree graphs. This finding reveals the geometry of metro networks
and tree graphs are in close proximity to one another. In $P$ space, the mean
Wasserstein distance between any pair of stations relates closely to the average
number of transfers. By ranking several key quantities transport concerned, we
obtain several ranking lists in which New York metro and Berlin metro
consistently top the first two spots.

\end{abstract}

\maketitle

Public transportation networks are crucial for cities. In mega cities metro
networks are major parts of transportation on which most city inhabitants rely
for daily mobility. It is therefore vital to evaluate the overall transport
performance of metro networks. Most related attempts focused on optimal routes
design which turned this problem into an engineering one aiming at the
optimization of multi-objective tasks. For instance, Mandl \cite{s1} considered three
separate problems, assignment of passengers to routes, assignment of vehicles to
routes and finding the vehicle routes in a given network. Yeung \textit{et al}
\cite{s2}
derived a simple and generic routing algorithm capable of considering all
individual path choices simultaneously, which has been tested on London
underground network. So far little attention has been paid to a comprehensive,
empirical study of performance of worldwide major metro networks (MNS) by
comparing the geometry features at the system level. This is exactly the main
target of our project.

In this report, we examined 28 major MNS worldwide (see supplementary materials
(SM) for the complete list and the sources of data) to study the relations
between geometric structures and transport properties. We present for MNS
empirical measurements of (i) fractal dimension, (ii) graph energy and graph
Laplacian energy, (iii) graph spectral distance, and (iv) the Wasserstein
distance. Further analysis of the above measures yields for MNS (I) the
power-law scaling between the energy (or Laplacian energy) per unit area and the
fractal dimension, with lower energy (costs per unit area) for higher
dimensioned networks, (II) the scaling between the Wasserstein distance and the
fractal dimension, which is in the vicinity of the theoretical curve based on
special regular tree graphs, and (III) the phylogenetic network which suggests
the geometric kinship among different MNS. These findings lead to several
rankings of key quantities transport related in which New York metro and Berlin
metro consistently top the first two spots.

Our data samples were taken from official websites of MNS. We considered both
unweighted metro networks (UMN) and weighted metro networks (WMN). Defined on
edges, the weight $w_{\mathrm{AB}}$  of a given edge AB is the number of different metro
lines which pass through stations A and B. Our key results were obtained in $L$
space, so was our main discussion. The sole quantity of interest in $P$ space is
the Wasserstein distance \cite{s3}. We present the values of topological quantities
such as the average degree $\langle k \rangle$, the clustering coefficient
$C$, and the average number of transfers $A_{\mathrm{T}}$ etc.~in Table S1 of the SM.
From this table one can have an immediate impression of the general picture of
the geometric patterns of the MNS. First,  $\langle k\rangle$ is close to 2 for 
most MNS, which
indicates the tree-like structures of MNS. Second, that the clustering
coefficient being close to 0 clearly indicates that cycles are rare in most MNS.
Third, for most MNS, the average strength $\langle s\rangle$ is nearly equal to
$\langle k\rangle$, which
implies weight does not play a significant role. Exceptions are Berlin, Hamburg,
Melbourne, Milan, New York, Seoul, Valencia and Washington. Fourth, the density
$\rho$ is very small, which marks the sparseness of the MNS.

It is straightforward that the MNS are fractal \cite{s4,s5,s6,s7}. Here we 
adopt the methods
of fractal dimension on networks (see SM). Denote the fractal dimension by $D$.
For UMN, $D$ ranges from 1.0895 (Montreal) to 1.8237 (New York). $D$ of Montreal is
close to the dimension of a line, which is exactly indicated by the shape of its
metro plan (see SM). $D$’s of UMN in Asia are all smaller than 1.5. Three European
cities have $D$’s larger than 1.5, with Berlin 1.6548, Paris 1.6005 and Milan
1.5524. When it comes to WMN, New York still tops with $D$ =1.8120, followed by
Milan with $D$ being 1.6801.

\begin{figure}[!htb]
	\begin{center}
		\includegraphics[width=0.5\textwidth]{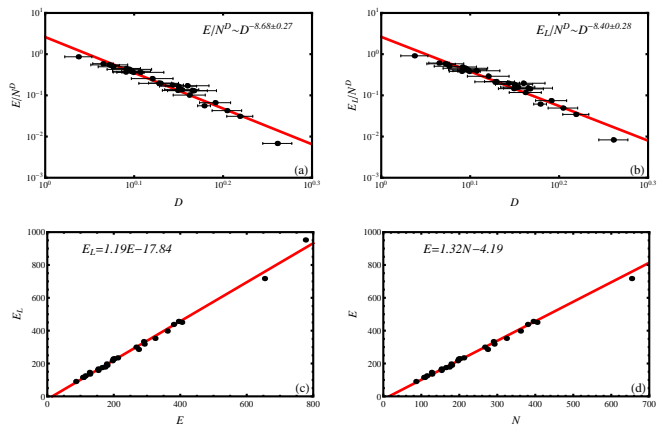}
	\end{center}
	\caption{For UMN there exists (a) the scaling between the rescaled
		energy$E/N^{D}$ and the fractal dimension $D$; (b) the scaling between the
		rescaled Laplacian energy $E_{L}/N^{D}$ and the fractal
		dimension $D$; (c) the linear
		relationship between the energy $E$  and the Laplacian energy
		$E_{L}$; (d) the linear
dependence of the energy $E$ on the size $N$.}
	\label{fig:1}
\end{figure}

To have a better understanding of the geometry of the MNS, the spectral graph
theory \cite{s8,s9,s10,s11,s12,s13,s14} and the graph energy theory \cite{s15,
s16,s17,s18} were applied to our data. The
former is an elegant application of differential geometry methods to discrete
spaces and mainly deals with the spectra of eigenvalues of matrices topology
concerned, such as adjacency matrix $A$ and Laplacian matrix $L$ (see SM). For a
graph $G$ with $N$ vertices, the energy $E$ is defined as
\begin{equation}
	E=\sum_{i}\abs{\lambda_{i}},
\end{equation}
where $\lambda_{i}(i=1,2,\ldots,N)$ is the
$i$-th eigenvalue of $A$ of $G$. The Laplacian energy $E_{L}$ of $G$ is defined
in a similar way as 
\begin{equation}
	E_{L}=\sum_{i}\abs{\mu_{i}-\langle k\rangle},
\end{equation}
where
$\mu_{i}(i=1,2,\ldots,N)$ is $i$-th eigenvalue of $L$ of $G$, and $\langle
k\rangle$ is the average vertex degree of $G$. We studied $E$ and $E_{L}$ for
both UMN (Fig.~\ref{fig:1}) and WMN (see SM). We only discussed the energy of
UMN as similar analysis can be applied to WMN.

The first and also the major finding is the striking scaling between the
rescaled energy $E/N^{D}$ and $D$
\begin{equation}
	E/N^{D}\sim D^{-8.68\pm 0.27},
\end{equation}
(see Fig.~\ref{fig:1}a), as well as the scaling between the rescaled Laplacian
energy $E_{L}/N^{D}$ and $D$
\begin{equation}
	E_{L}/N^{D}\sim D^{-8.40\pm 0.28},
\end{equation}
(see Fig.~\ref{fig:1}b). The second finding is that $E$ and $E_{L}$ are almost identical
(the two only differ slightly in scales) in describing the energy of a graph
(Fig.~\ref{fig:1}c). Therefore we only focus on one of them, say $E$. The third
finding is that both $E$ and $E_{L}$ scale as $N$ (Fig.~\ref{fig:1}d).

The second and third findings are not hard to interpret by checking the
definitions of $E$ and $E_{L}$. But what kind of information can be extracted
from the first finding? According to the graph energy theory, for a given
$N$, the star graph with the fewest connections $N-1$ uniquely has the smallest
energy $2\sqrt{N-1}$, and the complete graph owns the largest energy $2(N-1)$
(only exceeded by the very rare hyper-energetic graphs). MNS are neither star
graphs, nor complete graphs and must be in the middle between these two
extremes. For the MNS, more connections inevitably increase the construction
costs. Therefore, one can treat $E$ as energy cost. $N^{D}$ is analogous to the
area of a given metro with $N$ stations and fractal dimension $D$. Hence
$E/N^{D}$ can be viewed as the energy cost per unit area for a given graph with
parameters $E$, $N$, and $D$. For UMN, the highest value of $E/N^{D}$ is close
to 0.8635, for Montreal, and the lowest value of $E/N^{D}$ is 0.0068, for New
York. So by this criterion, the top 3 metros with the lowest energy costs per
unit area are New York, Berlin and Paris. And in Asian metros, Seoul and Tokyo
are the top 2. Delhi, Shenzhen, Taipei and Busan are on the bottom list with
higher energy costs per unit area. For WMN, the ranking list of $E/N^{D}$ is
nearly the same as the one for UMN.

To investigate the transport properties of the MNS, we employed the Wasserstein
distance \cite{s19,s20}, the key metric in optimal transport theory. Consider two
probability measures $\mu$ and $\nu$, with supports $\left\{x_{1},
x_{2},\ldots,x_{n} \right\}$ and $\left\{ y_{1}, y_{2},\ldots,y_{m} \right\}$
respectively. Then the Wasserstein distance between $\mu$ and $\nu$ is defined
as
\begin{equation}
	W_{1}(\mu,\nu)=\inf\limits_{\substack{
				\xi\\
				\sum_{j}\xi(x_{i},y_{j})=\mu(x_{i})\\
				\sum_{i}\xi(x_{i},y_{j})=\nu(y_{j})}}
				\sum_{i,j}c(x_{i},y_{j})\xi(x_{i},y_{j}),
\end{equation}
where $c(x,y)$ is the cost of transporting one unit of mass from point
$x$ to point $y$, and $\xi$ is a transfer plan between its margins $\mu$ and
$\nu$.

\begin{figure}[!htb]
	\begin{center}
		\includegraphics[width=0.5\textwidth]{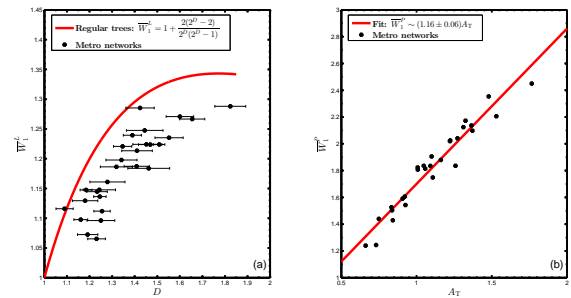}
	\end{center}
	\caption{The mean Wasserstein distance between connected stations
         $\oline{W}_{1}^{L}/D$ versus the fractal dimension $D$, in $L$ space. The solid curve indicates
the same measurement treated on homogeneous trees.  Data points being below the
curve implies that any given metro network has better transport features than
the regular tree with the same $D$. (b) The mean Wasserstein distance between any
two stations $\oline{W}_{1}^{P}/N_{\mathrm{L}}$ versus the average number of transfers
$A_{\mathrm{T}}$. The linear dependence
of the two implies that in $P$ space, $\oline{W}_{1}^{P}/N_{\mathrm{L}}$ simply specifies the convenience of metro
networks in regard of transfers needed.}
	\label{fig:2}
\end{figure}

We calculated the Wasserstein distances for the MNS in both $L$ space and $P$ space.
In $L$ space, we calculated the mean values of $W_{1}(x,y)$, denoted by
$\oline{W}_{1}^{L}$, between any pair
of stations which are directly connected. In principle, one can also calculate
the mean values of $W_{1}(x,y)$ between any two stations within the system. But as 
$W_{1}(x,y)$ is a
quantity for measuring the local geometric structure, $W_{1}(x,y)$ for nodes further
away
does not significantly relate to the transport properties. We found a scaling of
$\oline{W}_{1}^{L}$ versus $D$, with larger distance for higher dimension
(Fig.~\ref{fig:2}a). The largest
$\oline{W}_{1}^{L}$ comes from New York, being 1.2880, and the smallest one comes
from Melbourne, being 1.0653.

As for most MNS the mean degree $\langle k\rangle$ approximates 2, it is then
natural to relate MNS to tree graphs. For the sake of calculation, we consider a
special class of homogeneous tree graphs, in which $\oline{W}_{1}^{L}$,
$D$ and their relationship
\begin{equation}
\oline{W}_{1}^{L}=1+\frac{2(2^{D}-2)}{2^{D}(2^{D}-1)},
\end{equation}
can be exactly
obtained (see SM). In Fig.~\ref{fig:2}a, it can be seen that the theoretical curve for
regular trees of interest is close to the empirical results by using our data.
This observation confirms again the similarity in geometry between tree graphs
and the MNS. We also notice that for most MNS, the empirical data points are
below the theoretical values. This means that for any given $D$, the MNS have
better transport properties than the homogeneous trees constructed by having
smaller $\oline{W}_{1}^{L}$, the mean transport cost.

How do we understand the feature of the relationship between $\oline{W}_{1}^{L}$
and $D$? The
Wasserstein distance $W_{1}(x,y)$ for an edge $xy$ increases when $D$ becomes larger. But for
$\oline{W}_{1}^{L}$ on a homogeneous tree this behavior depends on how large
$D$ is. Recall by the formula (see SM), $W_{1}(x,y)$ takes constant values 1 on
the leaves and larger values $3-\frac{4}{k}$ on the other edges. Therefore, for 
$\oline{W}_{1}^{L}$, those non-leaf edges play the main role. However the
fraction of non-leaf edges in the whole edge set of a homogeneous tree will
decrease very quickly as $D$ or $k$ becomes very large. Hence when $D$ is large
enough, this decreasing of the fraction of non-leaf edges will balance the
increasing of the Wasserstein distance. Therefore, we observe a peak in the
theoretic curve.

In $P$ space, $\oline{W}_{1}^{P}$ was calculated between any pair of
stations. It is shown in Fig.~\ref{fig:2}b that $\oline{W}_{1}^{P}$ is
linearly proportional to the average transfer $A_{\mathrm{T}}$, which is rather
straightforward. As known in $P$ space, a single line is a complete subgraph.
Therefore, $W_{1}(x,y)$ between any two stations pertaining to the same line is simply 1
according to its definition. For two stations which do not belong to the same
line, the number of transfers shall concern in calculating $\oline{W}_{1}^{P}$.
Hence in $P$ space, $\oline{W}_{1}^{P}$ measures the convenience of transfer
a certain metro generally provides.
Larger  $\oline{W}_{1}^{P}$  states more transfers and smaller $\oline{W}_{1}^{P}$, less.

\begin{figure}[!htb]
	\begin{center}
		\includegraphics[width=0.5\textwidth]{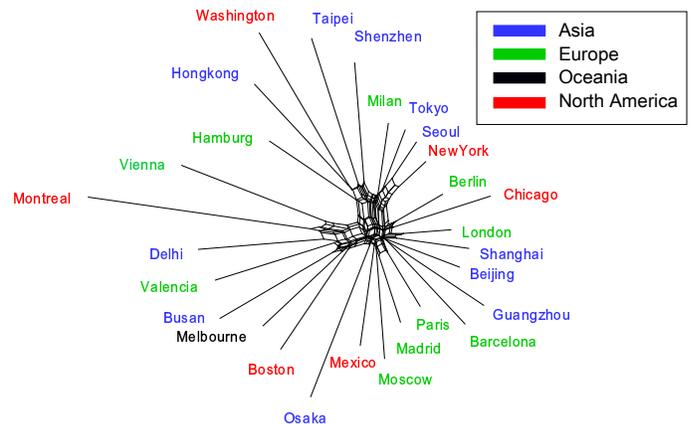}
	\end{center}
	\caption{The phylogenetic reconstruction of 28 metro networks, in which
	the graph distances among metro networks were visualized in a planar
graph. The clustering of certain metros may reveal the geometry kinship among
the communities.}
	\label{fig:3}
\end{figure}

Intuitively we know that the MNS are all somewhat unique in geometric
structures. But to quantitatively justify such differences we shall resort to
specialized techniques in graph spectra theory. We have obtained graph spectra
for all the MNS (see Fig. S10 of SM for examples). If one compares the spectra
of Berlin metro and Paris metro, the gap is big. But when it comes to the
spectra of Beijing metro and Berlin metro, these two are quite similar. To
capture the distinctions, the spectral distances based on the normalized
Laplacian were calculated and grouped, for the phylogenetic reconstruction
\cite{s21},
aiming at identifying the relationships in geometric structure of the MNS. The
quality of this reconstruction strongly depends on the distance matrix. The
phylogenetic network of the MNS is shown in Fig.~\ref{fig:3}. As seen, most of
the MNS in European cities are clustered, indicating the geometry kinship of
these networks. The MNS in Guangzhou, Shanghai, Shenzhen and Beijing are close
to each other, which manifests the proximity of designs of these metros. Fixing
a certain metro, say New York, one can roughly estimate from the phylogenetic
tree which of the rest metros is the closest, the next closest and the next to
the next closest, so on and so forth. The list then goes like this: New
York$\rightarrow$Milan$\rightarrow$Berlin\ldots

We now compare the key quantities for all MNS to obtain a series of ranking
lists. For any given metro network, if the ranking orders for different
quantities are consistent, then the ranking lists can be used to evaluate the
transport properties of the network of interest. For UMN, the ranking lists are
$E/N^{D}$, $A_{\mathrm{T}}/N_{\mathrm{L}}$, $\oline{W}_{1}^{P}/N_{\mathrm{L}}$,
$\oline{W}_{1}^{L}/D$ and $\rho$. Here $N_{\mathrm{L}}$ is the number of lines. For
WMN, the ranking lists are $E/N^{D}$ and $A_{\mathrm{T}}/N_{\mathrm{L}}$. To ensure the
consistency, all the lists are ranked from the highest to the lowest, regarding
the performance. By this convention, better performance corresponds to smaller
values of ranked quantities. $E/N^{D}$ is the energy cost, $A_{\mathrm{T}}/N_{\mathrm{L}}$ is the
transfer time cost, $\oline{W}_{1}^{P}/N_{\mathrm{L}}$ and $\oline{W}_{1}^{L}/D$
are the transport cost, and $\rho$ is the construction cost. From Table~\ref{tab:1},
New York and Berlin top the first two spots consistently. Melbourne, Milan, Paris
and Seoul are highly performed after New York and Berlin. Busan and Montreal are
on the bottom spots. Here we are certainly not suggesting that New York is the
best metro and Montreal is the worst. Rather, according to our criterion, New
York and Berlin provide some efficient structures which might be more beneficial
for transport. It is expected that this may shed some light on the planning of
future metros.

\noindent\textbf{Supplementary Materials:}\\
Materials and Methods\\
Figures S1-S10\\
Tables S1-S3\\
References[1-8]

\begin{center}
\begin{table*}[t]
	\centering
	{\scriptsize
	\hfill{}
	\begin{tabular}{lllll|ll}
\hline
\multicolumn{5}{c|}{\textbf{UMN}} & \multicolumn{2}{c}{\textbf{WMN}}\\
\hline
$E_{L}/N^{D}$ & $A_{\mathrm{T}}/N_{\mathrm{L}}$ & $\oline{W}_{1}^{P}/N_{\mathrm{L}}$ & $\oline{W}_{1}^{L}/D$ & $\rho$ & $E_{L}/N^{D}$ & $A_{\mathrm{T}}/N_{\mathrm{L}}$\\
\hline
0.0083 (New York)&0.0420 (New York)&0.0760 (New York)&0.7063 (New York)&0.0042 (New York)&0.0107 (New York)&0.0768 (New York)\\
0.0345 (Berlin)&0.0437 (Berlin)&0.0765 (Berlin)&0.7654 (Berlin)&0.0046 (Seoul)&0.0500 (Paris)&0.0857 (Berlin)\\
0.0492 (Paris)&0.0545 (Melbourne)&0.0907 (Melbourne)&0.7940 (Paris)&0.0076 (Berlin)&0.0510 (Milan)&0.0878 (Melbourne)\\
0.0613 (Seoul)&0.0576 (Seoul)&0.0945 (Seoul)&0.0796 (Milan)&0.0079 (Paris)&0.0546 (Berlin)&0.1246 (Hamburg)\\
0.0744 (Milan)&0.0606 (London)&0.0972 (London)&0.8095 (Moscow)&0.0082 (London)&0.0847 (Seoul)&0.1295 (Milan)\\
0.1171 (London)&0.0611 (Paris)&0.1059 (Paris)&0.8112 (Seoul)&0.0083 (Madrid)&0.1199 (Tokyo)&0.1415 (Seoul)\\
0.1434 (Barcelona)&0.0622 (Milan)&0.1067 (Milan)&0.8331 (Barcelona)&0.0088 (Shanghai)&0.128o (London)&0.1449 (Paris)\\
0.1455 (Madrid)& 0.0702 (Hamburg)&0.1191 (Hamburg)&0.8427 (Madrid)&0.0095 (Melbourne)&0.1366 (Barcelona)&0.1521 (London)\\
0.1494 (Moscow)& 0.0776 (Tokyo)&0.1328 (Madrid)&0.8431 (London)&0.0099 (Beijing)&0.1406 (Madrid)&0.1583 (Washington)\\
0.1582 (Tokyo)& 0.0819 (Madrid)&0.1390 (Tokyo)&0.8603 (Hamburg)&0.0105 (Milan)&0.1538 (Moscow)&0.1735 (Tokyo)\\
0.1850 (Hamburg)& 0.0840 (Moscow)&0.1411 (Moscow)&0.8636 (Osaka)&0.0112 (Tokyo)&0.1575 (Hamburg)&0.1824 (Chicago)\\
0.1968 (Osaka)&0.0873 (Shanghai)&0.1443 (Shanghai)&0.8653 (Melbourne)&0.0139 (Mexico)&0.1976 (Mexico)&0.2005 (Moscow)\\
0.1976 (Mexico)& 0.0923 (Barcelona)&0.1457 (Barcelona)&0.8766 (Boston)&0.0145 (Chicago)&0.2185 (Beijing)&0.2071 (Shanghai)\\
0.2132 (Shanghai)& 0.0930 (Chicago)&0.1569 (Beijing)&0.8857 (Valencia)&0.0148 (Valencia)&0.2222 (Osaka)&0.2120 (Osaka)\\
0.2167 (Beijing)&0.0986 (Beijing)&0.1671 (Chicago)&0.8918 (Mexico)&0.0148 (Delhi)&0.2318 (Shanghai)&0.2162 (Madrid)\\
0.2908 (Chicago)&0.1019 (Mexico)&0.1688 (Mexico)&0.8929 (Shanghai)&0.0151 (Hamburg)&0.3284 (Chicago)&0.2330 (Barcelona)\\
0.3797 (Valencia)&0.1041 (Osaka)&0.1879 (Valencia)&0.8993 (Chicago)&0.0156 (Moscow)&0.3955 (Vienna)&0.2591 (Mexico)\\
0.3820 (Melbourne)&0.1103 (Washington)&0.1885 (Taipei)&0.9001 (Delhi)&0.0164 (Guangzhou)&0.4110 (Guangzhou)&0.2596 (Valencia)\\
0.3954 (Vienna)&0.1162 (Valencia)&0.1907 (Boston)&0.9027 (Tokyo)&0.0164 (Barcelona)&0.4192 (Melbourne)&0.2971 (Beijing)\\
0.4110 (Guangzhou)& 0.1246 (Boston)&0.1908 (Osaka)&0.9066 (Beijing)&0.0172 (Busan)&0.4709 (Boston)&0.2998 (Shenzhen)\\
0.4194 (Boston)& 0.1358 (Taipei)&0.2066 (Washington)&0.9074 (Vienna)&0.0178 (Boston)&0.5087 (Delhi)&0.3588 (Boston)\\
0.4545 (Washington)& 0.1499 (Shenzhen)&0.2206 (HongKong)&0.9120 (Guangzhou)&0.0183 (Shenzhen)&0.5391 (Hong Kong)&0.4215 (Vienna)\\
0.4883 (Hong Kong)&0.1516 (Guangzhou)&0.2295 (Delhi)&0.9230 (Washington)&0.0204 (Taipei)&0.5691 (Shenzhen)&0.4245 (Montreal)\\
0.5086 (Delhi)&0.1530 (HongKong)&0.2374 (Guangzhou)&0.9279 (Hong Kong)&0.0240 (Vienna)&0.5837 (Taipei)&0.4327 (Guangzhou)\\
0.5691 (Shenzhen)&0.1573 (Delhi)&0.2879 (Shenzhen)&0.9451 (Busan)&0.0241 (Osaka)&0.6086 (Busan)&0.4480 (Delhi)\\
0.5837 (Taipei)&0.1815 (Vienna)&0.3111 (Montreal)&0.9568 (Taipei)&0.0241 (Washington)&0.6741 (Valencia)&0.4750 (Taipei)\\
0.6086 (Busan)&0.1827 (Montreal)&0.3181 (Vienna)&0.9686 (Shenzhen)&0.0250 (HongKong)&0.8904 (Washington)&0.4923 (HongKong)\\
0.9114 (Montreal)&0.1845 (Busan)&0.3213 (Busan)&1.0242 (Montreal)&0.0303 (Montreal)&0.9114 (Montreal)&0.6610 (Busan)\\
\hline
	\end{tabular}}
	\hfill{}
	\caption{Ranking of key quantities for UMN (unweighted metro networks) and WMN (weighted
		metro networks). $E_{L}/N^{D}$: the average Laplacian energy per area; $A_{\mathrm{T}}/N_{\mathrm{L}}$: the average number of
	transfers per line; $\oline{W}_{1}^{L}/D$: the ratio of mean Wasserstein distance in $L$ space to fractal dimension
	$D$; $\oline{W}_{1}^{P}/N_{\mathrm{L}}$: the ratio of mean Wasserstein distance to the number of lines $N_{\mathrm{L}}$; $\rho$: 
	the density of networks.}
	\label{tab:1}
\end{table*}
\end{center}

\textbf{Acknowledgements:} The source of data presented in this paper can be retrieved in the supplementary materials. W.~L.~would like to thank Prof.~Jost for the
hospitality during his stay in Leipzig where part of this work was done. This work was partially supported by the National Natural Science Foundation of China under 
grant No.~10975057, the Programme of Introducing Talents of Discipline to Universities under Grant No.~B08033, and Engineering And Physical Sciences Research Council
under grant No.~EP/K016687/1. The authors declare no competing financial interests.

\end{document}